\documentclass [6pt,a4paper]{article}
\usepackage{bbm}
\usepackage{amsfonts}
\usepackage{mathrsfs}
\usepackage {amssymb}
\usepackage {amsmath}
\usepackage{amsthm}
\usepackage{latexsym}
\usepackage[english]{babel}
\def\dse#1{\vskip 0.6cm\noindent
        {\large\bf #1}
        \vskip 0.4cm}
\def\dse#1{\vskip 0.6cm\noindent
        {\large\bf #1}
        \vskip 0.4cm}
\usepackage{lastpage}
\usepackage{epsfig}

\begin{document}
\begin{center}
\textbf{\large{New quantum MDS codes derived from constacyclic codes}}\footnote { E-mail addresses:
liqiwangg@163.com(L.Wang), zhushixin@hfut.edu.cn(S.Zhu).}
\end{center}

\begin{center}
{Liqi Wang, Shixin Zhu}
\end{center}

\begin{center}
\textit{School of Mathematics, Hefei University of
Technology Hefei 230009, Anhui, P.R.China}
\end{center}

\noindent\textbf{Abstract} Quantum maximal-distance-separable (MDS) codes form an important class of quantum codes. It is very hard to construct quantum MDS codes with relatively large minimum distance. In this paper, based on classical constacyclic codes, we construct two classes of quantum MDS codes with parameters $$[[\lambda(q-1),\lambda(q-1)-2d+2,d]]_q$$ where $2\leq d\leq (q+1)/2+\lambda-1$, and $q+1=\lambda r$ with $r$ even, and $$[[\lambda(q-1),\lambda(q-1)-2d+2,d]]_q$$ where $2\leq d\leq (q+1)/2+\lambda/2-1$, and $q+1=\lambda r$ with $r$ odd. The quantum MDS codes exhibited here have parameters better than the ones available in the literature.\\


\dse{1~~Introduction}

Quantum error-correcting codes play an important role in both quantum communication and quantum computation. It has experienced a great progress since the establishment of the connections between quantum codes and classical codes (see \cite{CRSS98}). It was shown that the construction of quantum codes can be reduced to that of classical linear error-correcting codes with certain self-orthogonality properties.

Let $q$ be a prime power. A $q$-ary quantum code $Q$ of length $n$ and size $K$ is a
$K$-dimensional subspace of a $q^n$-dimensional Hilbert space
$\mathbb{H}=\mathbb{C}^{q^n}=\mathbb{C}^q\otimes\cdots\otimes\mathbb{C}^q.$
The error correction and deletion capabilities of a quantum error-correcting code are the most crucial aspects of the code. If a quantum code has minimum distance $d$, then it can detect any $d-1$ and correct any $\lfloor(d-1)/2\rfloor$ quantum errors. Let $k=log_qK$. We use $[[n,k,d]]_q$ to denote a $q$-ary quantum code of length $n$ with size $q^k$ and minimum distance $d$. One of the principal problems in quantum error-correction is to construct quantum codes with the best possible minimum distance. It is well known that quantum codes with parameters $[[n,k,d]]_q$ must satisfy the quantum Singleton bound: $k\leq n-2d+2$ (see \cite{KKKS06} and \cite{KL97}). A quantum code achieving this bound is called a quantum maximum-distance-separable (MDS) code. Quantum MDS codes are the most useful and interesting quantum codes in quantum error correction.

In recent years, constructing quantum MDS codes has become one of the central topics for quantum codes. Several families of quantum MDS codes have been constructed (see \cite{BE00},\cite{F02}, \cite{GBR04}, \cite{GRB04}, \cite{HTZCYO08}, \cite{JLLX10}, \cite{JX14}, \cite{G11}, \cite{LXW08}, \cite{LX10}). As we know, if the classical MDS conjecture holds, the length of nontrivial $q$-ary quantum MDS codes cannot exceed $q^2+1$ (see \cite{KKKS06}). The problem of constructing quantum MDS codes with $n\leq q+1$ has been completely solved (see \cite{GBR04} and \cite{GRB04}). Many quantum MDS codes of length between $q+1$ and $q^2+1$ have also been constructed (see \cite{BE00}, \cite{JLLX10}, \cite{JX14}, \cite{G11}, \cite{LMPZ96}, \cite{LXW08}). Although so, there are still a lot of quantum MDS codes difficult to be constructed. It is a great challenge to construct new quantum MDS codes and a even more challenge to construct quantum MDS codes with relatively large minimum distance.

As mentioned in \cite{JX14}, except for some sparse lengths, almost all known $q$-ary quantum MDS codes have minimum distance less than or equal to $q/2+1$. Recently, Kai and Zhu constructed two new classes of quantum MDS codes based on classical negacyclic codes (see \cite{KZ13}) and six new classes of quantum MDS codes based on classical constacyclic codes (see \cite{KZ14}). These codes have minimum distance larger than $q/2+1$ in general. Two classes of the quantum MDS codes constructed in \cite{KZ14} are
\begin{description}
\item 1. $[[\lambda(q-1),\lambda(q-1)-2d+2,d]]_q$, where $\lambda=(q+1)/2$ and $2\leq d\leq q$, 
\item 2. $[[\lambda(q-1),\lambda(q-1)-2d+2,d]]_q$, where $2\leq d\leq (q+1)/2$ and $\lambda=(q+1)/r$ with even $r\neq2$.
\end{description}
We extend these two classes of quantum codes and obtain our first class of quantum MDS codes with parameters $[[\lambda(q-1),\lambda(q-1)-2d+2,d]]_q,$ where $2\leq d\leq (q+1)/2+\lambda-1$ and $q+1=\lambda r$ with $r$ even. This class of quantum MDS codes has larger minimum distance than theirs under the case even $r\neq2$. In particular, the obtained quantum codes have minimum distance bigger than $q/2+1$. Furthermore, we consider the case $r$ is an odd divisor of $q+1$ and get our second class of quantum MDS codes with parameters $[[\lambda(q-1),\lambda(q-1)-2d+2,d]]_q$, where $2\leq d\leq (q+1)/2+\lambda/2-1$ and $q+1=\lambda r$. It is new in sense that its parameters are not covered by the codes available in the literature, and it has minimum distance bigger than $q/2+1$ when $q\geq 11.$

This paper structured as follows. Section II presents a review of classical constacyclic code. In Section III, two classes of quantum MDS codes are derived by using Hermitian construction. Section IV concludes the paper.

\dse{2~~Review of constacyclic codes}

Let $\mathbb{F}_{q^2}$ be the Galois field with
$q^2$ elements, where $q$ is a prime power. A $q^2$-ary linear
code $C$ of length $n$ is a nonempty subspace of
$\mathbb{F}_{q^2}^n$. A $q^2$-ary linear code $C$ of length $n$ is called $\eta$-constacyclic if it is invariant under the $\eta$-constacyclic
shift of $\mathbb{F}_{q^2}^n$:
$$(c_0,c_1,\ldots,c_{n-1})\rightarrow(\eta c_{n-1},c_0,\ldots,c_{n-2}),$$
where $\eta$ is a nonzero element of $\mathbb{F}_{q^2}$.
Each codeword $\textbf{c}=(c_0,c_1,\ldots,c_{n-1})$ is customarily identified
with its polynomial representation
$c(x)=c_0+c_1x+\cdots+c_{n-1}x^{n-1}$, and the code $C$ is in turn
identified with the set of all polynomial representations of its
codewords. Then in the ring $\frac{\mathbb{F}_{q^2}[x]}{\langle
x^n-\eta\rangle},$ $xc(x)$ corresponds to a $\eta$-constacyclic shift of $c(x)$. It
is well known that a linear code $C$ of length $n$ over
$\mathbb{F}_{q^2}$ is $\eta$-constacyclic if and only if $C$ is an ideal of the
quotient ring $\frac{\mathbb{F}_{q^2}[x]}{\langle x^n-\eta\rangle}.$
Moreover, $\frac{\mathbb{F}_{q^2}[x]}{\langle x^n-\eta\rangle}$ is a
principal ideal ring, whose ideals are generated by monic factors of
$x^n-\eta$, i.e., $C=\langle f(x)\rangle$ and $f(x)|(x^n-\eta)$.

Given two vectors $\textbf{x}=(x_0, x_1,\ldots,x_{n-1})$, and $\textbf{y}=(y_0, y_1,\ldots,y_{n-1})\in\mathbb{F}_{q^2}^n$, their Hermitian inner product is defined as $$\langle \textbf{x}, \textbf{y}\rangle=x_0\bar{y}_0+x_1\bar{y}_1+\cdots+x_{n-1}\bar{y}_{n-1}\in\mathbb{F}_{q^2},$$
where $\bar{y}_i=y_i^q$.
The vectors $\textbf{x}$ and $\textbf{y}$ are called orthogonal with respect to the Hermitian inner product if $\langle \textbf{x}, \textbf{y}\rangle=0.$ For a $q^2$-ary linear code $C$ of length $n$, the Hermitian dual code of $C$ is defined as $$C^{\bot_H}=\{\textbf{x}\in\mathbb{F}_{q^2}^n|\langle\textbf{x}, \textbf{y}\rangle=0\ for\ all\  \textbf{y}\in C\}.$$
A linear code $C$ of length $n$ over $\mathbb{F}_{q^2}$ is called Hermitian self-orthogonal if $C\subseteq C^{\bot_H}$, and
it is called Hermitian self-dual if $C=C^{\bot_H}.$

We assume $\gcd(q, n)=1$. Let $\delta$ be a primitive $rn$th root of unity in some extension field of $\mathbb{F}_{q^2}$ such that $\delta^n=\eta$. Let $\xi=\delta^r,$ then $\xi$ is a primitive $n$th root of unity. Hence,  $$x^n-\eta=\prod_{i=0}^{n-1}(x-\delta\xi^i)=\prod_{i=0}^{n-1}(x-\delta^{1+ir}).$$

Let $\Omega=\{1+ir|0\leq i\leq n-1\}.$ For each $j\in\Omega,$ let $C_j$ be the $q^2$-cyclotomic coset modulo $rn$ containing $j$. Let $C$ be an $\eta$-constacyclic code of length $n$ over $\mathbb{F}_{q^2}$ with generator polynomial $g(x)$. Then the set $Z=\{j\in\Omega|g(\delta^j)=0\}$ is called the defining set of $C$. It is clearly to see the defining set of $C$ is a union of some $q^2$-cyclotomic cosets modulo $rn$ and $dim(C)=n-|Z|$. It is also easily to see $C^{\bot_H}$ has defining set $Z^{\bot_H}=\{z\in\Omega|-qz \mod\ rn\not\in Z\}$(See Ref. \cite{KZ14}).\\

Similar to cyclic codes, there exists the following BCH bound for constacyclic code (see [2, Theorem 2.2] or [15, Lemma 4]).\\

\noindent\textbf{Theorem 2.1} \emph{ (The BCH bound for constacyclic codes) Assume that $\gcd(q, n)=1$. Let $C=\langle g(x)\rangle$ be an $\eta$-constacyclic code of length $n$ over $\mathbb{F}_{q^2}$ with the roots $\{\delta^{1+ir}|0\leq i\leq d-2\}$, where $\delta$ is a primitive $rn$th root of unity. Then the minimum distance of $C$ is at least $d$.}\\

The following result presents a criterion to determine whether or not a given $q^2$-ary $\eta$-constacyclic code is dual-containing (see [12, Lemma 2.2]).\\

\noindent\textbf{Lemma 2.2} \emph{ Let $r$ be a positive divisor of $q+1$ and  $\eta\in\mathbb{F}_{q^2}^*$ be of order $r$. Let $C$ be an $\eta$-constacyclic code of length $n$ over $\mathbb{F}_{q^2}$ with defining set $Z\subseteq \Omega$, then $C$ contains its Hermitian dual code if and only if $Z\bigcap Z^{-q}=\emptyset,$ where $Z^{-q}=\{-qz\ \mod\ rn|z\in Z\}$.}

\dse{3~~Codes Construction}

Let $r=(q+1)/gcd(v, q+1)$ and $q$ be an odd prime power. In the next two parts, we give the construction of quantum MDS codes due to the case $r$ is even or odd by using Hermitian construction. First, we recall the Hermitian quantum code construction:\\

\noindent\textbf{Lemma 3.1 \cite{AK01}} \emph{ If $C$ is a $q^2$-ary $[n,k,d]$-linear code such that $C^{\bot_{H}}\subseteq C$, then there exists a $[[n, 2k-n, \geq d]]_q$ quantum code. }\\

A. Length $n=\lambda(q-1)$ with $\lambda$ a divisor of $q+1$ and $r$ even\\

Let $r=(q+1)/gcd(v, q+1)$ be even, for some $v\in\{1,2,\ldots,q\}$. Let $\xi=\omega^{v(q-1)}$ and $\lambda=(q+1)/r$. Based on $\xi$-constacyclic codes, we first construct $q$-ary quantum MDS codes of length $\lambda(q-1)$. It is easy to see that the $q^2$-cyclotomic coset containing $1+jr+\frac{r-2}{2}(q+1)$ modulo $nr$ has only one element $1+jr+\frac{r-2}{2}(q+1)$, i.e., $C_{1+jr+\frac{r-2}{2}(q+1)}=\{1+jr+\frac{r-2}{2}(q+1)\}$ under the case $0\leq j\leq\frac{r+2}{2r}(q+1)$.\\

\noindent\textbf{Lemma 3.2 } \emph{ Let $r=(q+1)/\gcd(v, q+1)$ be even and $r\neq q+1$, for some $v\in\{1,2,\ldots,q\}$. Let $n=\lambda(q-1)$ with $\lambda=(q+1)/r$. Suppose that $C$ is an $\xi$-constacyclic code of length $n$ over $\mathbb{F}_{q^2}$ with defining set $Z=\bigcup_{j=1}^{\delta}C_{1+r(j-1)+\frac{r-2}{2}(q+1)}$, where $1\leq \delta\leq\frac{r+2}{2r}(q+1)-2$, then $C^{\perp_H}\subseteq C$.}\\

\noindent\textbf{Proof.} Suppose that $C$ does not contain its Hermitian dual code, then by Lemma 2.2, $Z\bigcap Z^{-q}\neq\emptyset$. Hence, there exist two integers $k,l\in\{1,2,\ldots,\frac{r+2}{2r}(q+1)-2\}$ such that $$1+r(k-1)+\frac{r-2}{2}(q+1)\equiv -[1+r(l-1)+\frac{r-2}{2}(q+1)]q\mod\ rn,$$
which is equivalent to
\begin{align}
k+ql-\lambda\equiv 0\mod\lambda(q-1). \label{eq:A}
\end{align}
Let $r=2s$ with some integer $s\geq 1$. Then $1\leq l\leq\lambda(s+1)-2$. We express $l$ in the form $l=u\lambda+v$, where $0\leq u\leq s$ and $0\leq v\leq\lambda-2$\ (except for the case $u=v=0$). We now consider it due to the following two cases.
\begin{description}
\item (1) $0\leq u\leq s$ and $1\leq v\leq\lambda-2.$ The congruence (1) yields $k+l+(q-1)v-\lambda\equiv 0\mod\ \lambda(q-1)$. Since $1\leq k,l\leq\frac{r+1}{2r}(q+1)-2<q-1,$ it follows that $k+l+(q-1)v-\lambda<(q-1)+(q-1)+(\lambda-2)(q-1)-\lambda=\lambda(q-2)$. This gives a contradiction.
\item (2) $1\leq u\leq s$ and $v=0.$ The congruence (1) yields $k+l-\lambda\equiv 0\mod\ \lambda(q-1)$. But $k+l-\lambda<2(q-1)-\lambda<2(q-1)$ and $\lambda>1$. This gives a contradiction.
\end{description}
This completes the proof.\qed\\

\noindent\textbf{Theorem 3.3} \emph{ Let $r$ be an even divisor of $q+1$ and $r\neq q+1$, and let $n=\lambda(q-1)$ with $\lambda=(q+1)/r$. Then there exist a $[[n,n-2d+2,d]]_q$ quantum MDS code, where $2\leq d\leq\frac{r+2}{2r}(q+1)-1.$}\\

\noindent\textbf{Proof.} Let $\xi=\omega^{\lambda(q-1)}$, where $\omega$ is a primitive element of $\mathbb{F}_{q^2}$. Let $C$ be the $\xi$-constacyclic code of length $n$ over $\mathbb{F}_{q^2}$ with defining set $Z=\bigcup_{j=1}^{\delta}C_{1+r(j-1)+\frac{r-2}{2}(q+1)}$, where $1\leq \delta\leq\frac{r+2}{2r}(q+1)-2$. It follows from Lemma 3.2 that $C$ contains its Hermitian dual code, and $dim(C)=n-\delta$. The BCH bound for constacyclic codes gives that the distance of $C$ is at least $\delta+1$. Hence, $C$ is a constacyclic code with parameters $[n, n-\delta, \geq\delta+1]_{q^2}$. Using the Hermitian construction, we obtain a $[[n, n-2\delta,\geq\delta+1]]_q$ quantum code. Combining the quantum Singleton bound yields a quantum code with parameters $[[n, n-2\delta,\delta+1]]_q$, which is the desired quantum MDS code.\qed\\

Taking $r=2$ in Theorem 3.3, we obtain a quantum MDS code with parameters $[[(q^2-1)/2,(q^2-1)/2-2d+2, d]]_q,$ where $2\leq d\leq q$, which is the quantum MDS codes got in [12, Theorem 3.2]. But when $r$ is an even divisor of $q+1$ and $r\neq 2$, there exists a quantum code in \cite{KZ14} with parameters  $[[\lambda(q-1),\lambda(q-1)-2d+2, d]]_q,$ where $2\leq d\leq (q+1)/2$. However, the construction in Theorem 3.3 produces some new quantum MDS codes with parameters  $[[\lambda(q-1),\lambda(q-1)-2d+2, d]]_q,$ where $(q+1)/2+1\leq d\leq(q+1)/2+\lambda-1$. These codes have much larger minimum distance in general. Hence, we unite the results of $r=2$ and $r\neq 2$ in \cite{KZ14} and more quantum MDS codes are obtained in our construction.\\

\noindent\textbf{Example 3.4} \emph{ Let $q=19$. Applying Theorem 3.3 produces four new quantum MDS codes with parameters $[[90,70,11]]_{19}$, $[[90,68,12]]_{19}$, $[[90,66,13]]_{19}$ and $[[90,64,14]]_{19}$.}\\

\noindent\textbf{Example 3.5} \emph{ Let $q=23$. Applying Theorem 3.3 produces some new quantum MDS codes in Table I.}

\begin{table}[htbp]
{\small
\begin{center}
{\small{\bf TABLE I}~~New quantum MDS codes\\}
\begin{tabular}{c c c c c c c c c c c c c c c c c c c}
\hline
 &  & $\lambda$ & & & $r$ &  & & $n$&  & &$[[n,k,d]]_{q}$& &  \\
\hline
 &  & 2& &  & 12 & &  &44& &  &$[[44,20,13]]_{23}$& & \\
 &  & 3& &  &8 & & &66 & & &$[[66,42,13]]_{23}$& &\\
 &  & 3& &  &8 & & &66 & & &$[[66,40,14]]_{23}$& &\\
 &  & 4& &  &6 & & &88 & & &$[[88,64,13]]_{23}$& &\\
 &  & 4& &  &6 & & &88 & & &$[[88,62,14]]_{23}$& &\\
 &  & 4& &  &6 & & &88 & & &$[[88,60,15]]_{23}$& &\\
 &  & 6& & &4& & &132& & &$[[132,108,13]]_{23}$& &\\
 &  & 6& & &4& & &132& & &$[[132,106,14]]_{23}$& &\\
 &  & 6& & &4& & &132& & &$[[132,104,15]]_{23}$& &\\
 &  & 6& & &4& & &132& & &$[[132,102,16]]_{23}$& &\\
 &  & 6& & &4& & &132& & &$[[132,100,17]]_{23}$& &\\
\hline
\end{tabular}
\end{center}}
\end{table}

B. Length $n=\lambda(q-1)$ with $\lambda$ a divisor of $q+1$ and $r$ odd\\

In \cite{KZ14}, the authors only consider the case $r$ is an even divisor of $q+1$, but how about the case $r$ is an odd divisor of $q+1$. In this part, we consider this problem and construct a new class of quantum MDS codes under such case.

Let $r=(q+1)/gcd(v, q+1)$ be odd, for some $v\in\{1,2,\ldots,q\}$. Let $\xi=\omega^{v(q-1)}$ and $\lambda=(q+1)/r$. Based on $\xi$-constacyclic codes, we construct $q$-ary quantum MDS codes of length $\lambda(q-1)$. It is easy to see that the $q^2$-cyclotomic coset containing $1+jr+\frac{r-1}{2}(q+1)$ modulo $nr$ has only one element $1+jr+\frac{r-1}{2}(q+1)$, i.e., $C_{1+jr+\frac{r-1}{2}(q+1)}=\{1+jr+\frac{r-1}{2}(q+1)\}$ under the case $0\leq j\leq\frac{r+1}{2r}(q+1)$.\\

\noindent\textbf{Lemma 3.6 } \emph{ Let $r=(q+1)/\gcd(v, q+1)$ be odd, for some $v\in\{1,2,\ldots,q\}$. Let $n=\lambda(q-1)$ with $\lambda=(q+1)/r$. Suppose that $C$ is an $\xi$-constacyclic code of length $n$ over $\mathbb{F}_{q^2}$ with defining set $Z=\bigcup_{j=1}^{\delta}C_{1+r(j-1)+\frac{r-1}{2}(q+1)}$, where $1\leq \delta\leq\frac{r+1}{2r}(q+1)-2$, then $C^{\perp_H}\subseteq C$.}\\

\noindent\textbf{Proof.} Suppose that $C$ does not contain its Hermitian dual code, then $Z\bigcap Z^{-q}\neq\emptyset$. Hence, there exist two integers $k,l\in\{1,2,\ldots,\frac{r+1}{2r}(q+1)-2\}$ such that $$1+r(k-1)+\frac{r-1}{2}(q+1)\equiv -[1+r(l-1)+\frac{r-1}{2}(q+1)]q\mod\ rn,$$
which is equivalent to
\begin{align}
k+ql\equiv 0\mod\lambda(q-1). \label{eq:A}
\end{align}
Let $r=2s+1$ with some integer $s\geq 1$. Then $1\leq l\leq\lambda(s+1)-2$. We express $l$ in the form $l=u\lambda+v$, where $0\leq u\leq s$ and $0\leq v\leq\lambda-2$\ (except for the case $u=v=0$). We now consider it due to the following two cases.
\begin{description}
\item (1) $0\leq u\leq s$ and $1\leq v\leq\lambda-2.$ The congruence (2) yields $k+l+(q-1)v\equiv 0\mod\ \lambda(q-1)$. Since $1\leq k,l\leq\frac{r+1}{2r}(q+1)-2<q-1,$ it follows that $k+l+(q-1)v<(q-1)+(q-1)+(\lambda-2)(q-1)=\lambda(q-1)$. This gives a contradiction.
\item (2) $1\leq u\leq s$ and $v=0.$ The congruence (2) yields $k+l\equiv 0\mod\ \lambda(q-1)$. But $k+l<2(q-1)$ and $\lambda>1$. This gives a contradiction.
\end{description}
This completes the proof.\qed\\

\noindent\textbf{Theorem 3.7} \emph{ Let $r$ be an odd divisor of $q+1$ and $n=\lambda(q-1)$ with $\lambda=(q+1)/r$. Then there exist a $[[n,n-2d+2,d]]_q$ quantum MDS code, where $2\leq d\leq\frac{r+1}{2r}(q+1)-1.$}\\

\noindent\textbf{Proof.} Let $\xi=\omega^{\lambda(q-1)}$, where $\omega$ is a primitive element of $\mathbb{F}_{q^2}$. Let $C$ be the $\xi$-constacyclic code of length $n$ over $\mathbb{F}_{q^2}$ with defining set $Z=\bigcup_{j=1}^{\delta}C_{1+r(j-1)+\frac{r-1}{2}(q+1)}$, where $1\leq \delta\leq\frac{r+1}{2r}(q+1)-2$. It follows from Lemma 3.6 that $C$ contains its Hermitian dual code. Note that $dim(C)=n-\delta$ and $d(C)\geq \delta+1$. So $C$ is a $\xi$-constacyclic code with parameters $[n, n-\delta, \geq\delta+1]_{q^2}$. Combining the Hermitian construction and the quantum Singleton bound, we obtain a $[[n, n-2\delta,\delta+1]]_q$ quantum code, which is the desired quantum MDS code.\qed\\

The construction in Theorem 3.7 produces some quantum MDS codes with parameters $[[\lambda(q-1),\lambda(q-1)-2d+2, d]]_q,$ where $(q+1)/2\leq d\leq(q+1)/2+\lambda/2-1$. These codes are new in the sense that their parameters are not covered in the literature and this class of quantum MDS codes has larger minimum distance than the known ones.\\

\noindent\textbf{Example 3.8} \emph{ Let $q=17$. Applying Theorem 3.7 produces four new quantum MDS codes in Table II.}
\begin{table}[htbp]
{\small
\begin{center}
{\small{\bf TABLE II}~~New quantum MDS codes\\}
\begin{tabular}{c c c c c c c c c c c c c c c c c c c}
\hline
 &  & $\lambda$ & & & $r$ &  & & $n$&  & &$[[n,k,d]]_{q}$& &  \\
\hline
 &  & 2& &  & 9 & &  &32& &  &$[[32,16,9]]_{17}$& & \\
 &  & 6& &  &3 & & &96 & & &$[[96,80,9]]_{17}$& &\\
 &  & 6& &  &3 & & &96 & & &$[[96,78,10]]_{17}$& &\\
 &  & 6& &  &3 & & &96 & & &$[[96,76,11]_{17}$& &\\
\hline
\end{tabular}
\end{center}}
\end{table}

\noindent\textbf{Example 3.9} \emph{ Let $q=29$. Applying Theorem 3.7 produces some new quantum MDS codes in Table III.}

\begin{table}[htbp]
{\small
\begin{center}
{\small{\bf TABLE III}~~New quantum MDS codes\\}
\begin{tabular}{c c c c c c c c c c c c c c c c c c c}
\hline
 &  & $\lambda$ & & & $r$ &  & & $n$&  & &$[[n,k,d]]_{q}$& &  \\
\hline
 &  & 2& &  & 15 & &  &56& &  &$[[56,28,15]]_{29}$& & \\
 &  & 6& &  &5 & & &168 & & &$[[168,140,15]]_{29}$& &\\
 &  & 6& &  &5 & & &168 & & &$[[168,138,16]]_{29}$& &\\
 &  & 6& &  &5 & & &168 & & &$[[168,136,17]]_{29}$& &\\
 &  & 10& &  &3 & & &280 & & &$[[280,252,15]]_{29}$& &\\
 &  & 10& &  &3 & & &280 & & &$[[280,250,16]]_{29}$& &\\
 &  & 10& &  &3 & & &280 & & &$[[280,248,17]]_{29}$& &\\
 &  & 10& &  &3 & & &280 & & &$[[280,246,18]]_{29}$& &\\
 &  & 10& &  &3 & & &280 & & &$[[280,244,19]]_{29}$& &\\
\hline
\end{tabular}
\end{center}}
\end{table}

\dse{4~~Conclusion}

We have constructed two classes of quantum MDS codes whose parameters are given by $[[\lambda(q-1),\lambda(q-1)-2d+2,d]]_q$, where $2\leq d\leq (q+1)/2+\lambda-1$ and $q+1=\lambda r$ with $r$ even, and $[[\lambda(q-1),\lambda(q-1)-2d+2,d]]_q$, where $2\leq d\leq (q+1)/2+\lambda/2-1$ and $q+1=\lambda r$ with $r$ odd. The first class of quantum MDS codes has much larger minimum distance than the known ones. The second class of quantum MDS codes is new in the sense that its parameters are different from all the known ones and they also have much larger minimum distances. It would be interesting to go on this line of study and more new quantum codes with good parameters may be constructed from classical constacyclic codes.

\end{document}